\documentclass[5p,twocolumn]{elsarticle}

\usepackage{amsmath,amssymb}
\newcommand{\eqdef}{\stackrel{\text{def}}{=}}
\newcommand{\n}{\nonumber \\}
\newcommand{\bm}{\boldsymbol}
\newcommand{\Romannumeral}[1]{\uppercase\expandafter{\romannumeral#1}}
\newcommand{\ignore}[1]{}

\allowdisplaybreaks[4]

\journal{Physics Letters B}

\begin{document}

\begin{frontmatter}

%%%%%%%%%%%%%%%%%%%%%%%%%%%%%%%%%%%%%%%%%%%%%%%%%%%%%%%%%%%%
%                                                          %
%  Title page                                              %
%                                                          %
%%%%%%%%%%%%%%%%%%%%%%%%%%%%%%%%%%%%%%%%%%%%%%%%%%%%%%%%%%%%
\title{Another set of infinitely many exceptional ($X_{\ell}$)
Laguerre polynomials}

\author[SO]{Satoru Odake\corref{cSO}}
\ead{odake@azusa.shinshu-u.ac.jp}
\author[RS]{Ryu Sasaki}
\address[SO]{Department of Physics, Shinshu University,
     Matsumoto 390-8621, Japan}
\address[RS]{Yukawa Institute for Theoretical Physics,
     Kyoto University, Kyoto 606-8502, Japan}
\cortext[cSO]{Corresponding author.}

\begin{abstract}
We present a new set of infinitely many shape invariant potentials and
the corresponding exceptional ($X_{\ell}$) Laguerre polynomials.
They are to supplement the recently derived two sets of infinitely
many shape invariant thus exactly solvable potentials in one dimensional
quantum mechanics and the corresponding $X_{\ell}$ Laguerre and Jacobi
polynomials (Odake and Sasaki, Phys. Lett. {\bf B679} (2009) 414-417).
The new $X_{\ell}$ Laguerre polynomials and the potentials are obtained
by a simple limiting procedure from the known $X_{\ell}$ Jacobi polynomials
and the potentials, whereas the known $X_{\ell}$ Laguerre polynomials and
the potentials are obtained in the same manner from the mirror image of
the known $X_{\ell}$ Jacobi polynomials and the potentials.
\end{abstract}

\begin{keyword}
shape invariance \sep orthogonal polynomials
\PACS 03.65.-w \sep 03.65.Ca \sep 03.65.Fd \sep 03.65.Ge \sep 02.30.Ik
\sep 02.30.Gp
\end{keyword}

%%%%%%%%%%%%%%%%%%%%%%%%%%%%%%%%%%%%%%%%%%%
% PACS 2006
% 02.       Mathematical methods in physics
%   02.20.Uw  Quantum Groups
%   02.30.Gp  Special functions
%   02.30.Ik  Integrable systems
% 03.       Quantum mechanics, field theories, and special relativity
%   03.65.-w  Quantum mechanics
%   03.65.Ca  Formalism
%   03.65.Fd  Algebraic methods
%   03.65.Ge  Solutions of wave equations: bound states
% 45.       Classical mechanics of discrete systems
%   45.20.-d  Formalisms in classical mechanics
%%%%%%%%%%%%%%%%%%%%%%%%%%%%%%%%%%%%%%%%%%%

%%%%%%%%%%%%%%%%%%%%%%%%%%%%%%%%%%%%%%%%%%%
% PACS 2008
% 02.00.00 Mathematical methods in physics
%   02.30.Gp Special functions
%   02.30.Ik Integrable systems
% 03.00.00 Quantum mechanics, field theories, and special relativity
%   03.65.-w Quantum mechanics
%   03.65.Ca Formalism
%   03.65.Db Functional analytical methods
%   03.65.Fd Algebraic methods
%   03.65.Ge Solutions of wave equations: bound states
% 45.00.00 Classical mechanics of discrete systems
%   45.20.-d Formalisms in classical mechanics
%%%%%%%%%%%%%%%%%%%%%%%%%%%%%%%%%%%%%%%%%%%

\end{frontmatter}

%%%%%%%%%%%%%%%%%%%%%%%%%%%%%%%%%%%%%%%%%%%%%%%%%%%%%%%%%%%%%%%
%                                                             %
%  1. Introduction                                            %
%                                                             %
%%%%%%%%%%%%%%%%%%%%%%%%%%%%%%%%%%%%%%%%%%%%%%%%%%%%%%%%%%%%%%%
\section{Introduction}
\label{intro}
\setcounter{equation}{0}

For a long time it was believed, due to Bochner's theorem \cite{bochner},
that among countless orthogonal polynomials, only the
{\em classical orthogonal polynomials\/}, the Hermite, Laguerre, Jacobi
and Bessel polynomials, satisfy second order differential equations.
In 2008 the notion of the {\em exceptional\/} ($X_{\ell}$) orthogonal
polynomials was introduced by Gomez-Ullate et al \cite{gomez} in the
framework of Sturm-Liouville theory.
These orthogonal polynomials are exceptional in the sense that they
start at degree $\ell$ ($\ell\ge1$) instead of degree 0 constant term,
thus avoiding restrictions of Bochner's theorem and they satisfy second
order differential equations. They constructed the lowest examples,
the $X_1$ Laguerre and $X_1$ Jacobi polynomials explicitly.
Reformulation \cite{quesne} within the framework of one dimensional
quantum mechanics and shape invariant potentials \cite{genden} followed.
Two sets of infinitely many shape invariant potentials, the deformed
radial oscillator potentials and the deformed trigonometric/hyperbolic
Darboux-P\"oschl-Teller (DPT) potentials, and the corresponding $X_{\ell}$
Laguerre and Jacobi polynomials ($\ell=1,2,\ldots,\infty$) were
presented by the present authors \cite{os16} in June 2009.
The $\ell=1$ examples are the same as those given by Gomez-Ullate et al.
\cite{gomez} and Quesne \cite{quesne}.
Shape invariance of the $\ell$-th members of the exactly solvable
potentials are attributed to new polynomial identities of degree $3\ell$
involving cubic products of the Laguerre or Jacobi polynomials and
are proved elementarily in \cite{os18}.

In this Letter, we present a new set of infinitely many shape invariant
potentials, deformed radial oscillator potentials and the corresponding
$X_{\ell}$ Laguerre polynomials.
They are obtained from the known \cite{os16} deformed trigonometric DPT
potential and the known $X_{\ell}$ Jacobi polynomials in a certain limit.
On the other hand the deformed radial oscillator potential in \cite{os16}
and the corresponding $X_{\ell}$ Laguerre polynomials are shown to be derived
in the same way from the `mirror image' of those deformed trigonometric
DPT potential and the corresponding $X_{\ell}$ Jacobi polynomials given in
\cite{os16}.
The first ($\ell=1$) members of the two exceptional Laguerre polynomials
are identical and the $X_1$ Jacobi polynomials and their mirror images are
also identical. This is one of the reasons why the new $X_{\ell}$ polynomials
were not discovered earlier.
The second ($\ell=2$) member of the new exceptional Laguerre polynomials
and its shape invariant potential are the same as those found by Quesne
\cite{quesne2}. They were called type \Romannumeral{2} exceptional Laguerre
polynomials and were discussed in some detail in \cite{tanaka}.
In this connection so-called type \Romannumeral{3} Laguerre and Jacobi
solutions were discussed by Quesne \cite{quesne2}. The nature of the two
type \Romannumeral{3} solutions in \cite{quesne2} will be explained in
the final section.

%\bigskip
This Letter is organised as follows.
In section two we briefly recapitulate the scheme of deformation of shape
invariant potentials in terms of a polynomial eigenfunction of degree $\ell$.
Various results of \cite{os16} are listed for comparison with the results
to be derived.
In section three the new set of infinitely many deformed oscillator
potentials and the corresponding $X_{\ell}$ Laguerre polynomials are derived
by a simple limiting procedure from the known deformed trigonometric DPT
potentials and the corresponding $X_{\ell}$ Jacobi polynomials in \cite{os16}.
Various limiting formulas are also displayed.
The final section is for a summary and comments.

%%%%%%%%%%%%%%%%%%%%%%%%%%%%%%%%%%%%%%%%%%%%%%%%%%%%%%%%%%%%%%%
%                                                             %
%  2. Deformed Shape Invariant Potentials                     %
%                                                             %
%%%%%%%%%%%%%%%%%%%%%%%%%%%%%%%%%%%%%%%%%%%%%%%%%%%%%%%%%%%%%%%
\section{Deformed Shape Invariant Potentials}
\label{deformedpot}

Here we follow the notation of \cite{os16} and recapitulate its main
results for comparison with the new results to be derived in the next
section. Exactly solvable deformation of the radial oscillator potential
and the trigonometric DPT potential is most easily achieved at the
prepotential level ($\ell=1,2,\ldots$):
\begin{equation}
  w_{\ell}(x;\bm{\lambda})\eqdef w_0(x;\bm{\lambda}+\ell\bm{\delta})
  +\log\frac{\xi_{\ell}(\eta(x);\bm{\lambda}+\bm{\delta})}
  {\xi_{\ell}(\eta(x);\bm{\lambda})},
  \label{wl}
\end{equation}
in which $w_0$ is the undeformed prepotential and $\xi_{\ell}$ is related
to the $\ell$-th eigenpolynomial of the undeformed system, to be explained
shortly. For the explanation of the sinusoidal coordinate $\eta(x)$ and
the sets of parameters $\bm{\lambda}$ and $\bm{\delta}$, see \cite{os16}.
These deformed prepotentials satisfy the shape invariance condition,
\begin{align}
  &\bigl(\partial_xw_{\ell}(x;\bm{\lambda})\bigr)^2
  -\partial_x^2w_{\ell}(x;\bm{\lambda})\n
  =&\,\bigl(\partial_xw_{\ell}(x;\bm{\lambda}+\bm{\delta})\bigr)^2
  +\partial_x^2w_{\ell}(x;\bm{\lambda}+\bm{\delta})
  +\mathcal{E}_1(\bm{\lambda}+\ell\bm{\delta}).
  \label{wlshapeinv}
\end{align}
The Hamiltonian and the other quantities are defined as:
\begin{align}
  &\mathcal{H}_{\ell}(x;\bm{\lambda})\eqdef
  \mathcal{A}_{\ell}(\bm{\lambda})^{\dagger}\mathcal{A}_{\ell}(\bm{\lambda})
  \!=p^2+U_{\ell}(x;\bm{\lambda}),\ p=-i\partial_x,
  \label{Hldef}\\
  &\mathcal{A}_{\ell}(\bm{\lambda})\eqdef\partial_x
  \!-\!\partial_xw_{\ell}(x;\bm{\lambda}),
  \ \mathcal{A}_{\ell}(\bm{\lambda})^{\dagger}
  \!=\!-\partial_x\!-\!\partial_xw_{\ell}(x;\bm{\lambda}),\!\!\\
  &U_{\ell}(x;\bm{\lambda})\eqdef
  \bigl(\partial_xw_{\ell}(x;\bm{\lambda})\bigr)^2
  +\partial_x^2w_{\ell}(x;\bm{\lambda}),
  \label{Udef}\\
  &\mathcal{H}_{\ell}(x;\bm{\lambda})\phi_{\ell,n}(x;\bm{\lambda})
  =\mathcal{E}_n(\bm{\lambda}+\ell\bm{\delta})\phi_{\ell,n}(x;\bm{\lambda}),\\
  &\phi_{\ell,n}(x;\bm{\lambda})
  =\psi_{\ell}(x;\bm{\lambda})P_{\ell,n}\bigl(\eta(x);\bm{\lambda}\bigr),\\
  &\psi_{\ell}(x;\bm{\lambda})\eqdef
  \frac{e^{w_0(x;\bm{\lambda}+\ell\bm{\delta})}}
  {\xi_{\ell}(\eta(x);\bm{\lambda})}.
  \label{genmeasure}
\end{align}
The polynomial eigenfunctions, {\em i.e.}, the {\em exceptional orthogonal
polynomials\/} $P_{\ell,n}(\eta;\bm{\lambda})$ form the
complete basis of the Hilbert space and satisfy the orthogonality:
\begin{equation}
  \int_{x_1}^{x_2}\!\!\psi_{\ell}(x;\bm{\lambda})^2
  P_{\ell,n}\bigl(\eta(x);\bm{\lambda}\bigr)
  P_{\ell,m}\bigl(\eta(x);\bm{\lambda}\bigr)dx
  =h_{\ell,n}(\bm{\lambda})\delta_{nm}.
  \label{hln}
\end{equation}
Here we list the explicit forms of various quantities.
We attach superscripts $\text{L}$ and $\text{J}$ for the quantities
related to the radial oscillator potential and the trigonometric DPT
potential, respectively. We also attach superscripts $1$ and $2$ to
distinguish those derived in the previous paper \cite{os16} and those
new quantities to be introduced in the next section, respectively.
\\[2pt]
{\bf radial oscillator} undeformed ($\ell=0$) case:
\begin{align}
  &\bm{\lambda}^{\text{L}}\eqdef g,\ \ \bm{\delta}^{\text{L}}=1,\ \ g>0,
  \label{raddat1}\\
  &\mathcal{E}_n^{\text{L}}(\bm{\lambda})=4n,
  \ \ \eta^{\text{L}}(x)\eqdef x^2,\ \ 0<x<\infty,
  \label{ratdomain}\\
  &\phi_0^{\text{L}}(x;\bm{\lambda})\eqdef e^{-\tfrac{x^2}{2}}x^g
  \Leftrightarrow
  w_0^{\text{L}}(x;\bm{\lambda})\eqdef-\frac{x^2}{2}+g\log x,\\
  &P_n^{\text{L}}(x;\bm{\lambda})\eqdef L_n^{(g-\tfrac{1}{2})}(x),\\
  &h_n^{\text{L}}(\bm{\lambda})=\frac{1}{2\,n!}\,\Gamma(n+g+\tfrac12).
\end{align}
\\[1pt]
$\ell$-th deformed {\bf radial oscillator}:
\begin{align}
  &\xi_{\ell}^{\text{L1}}(x;\bm{\lambda})\eqdef
  L_{\ell}^{(g+\ell-\tfrac{3}{2})}(-x),
  \label{L1xi}\\
  &P_{\ell,n}^{\text{L1}}(x;\bm{\lambda})\eqdef\xi_{\ell}^{\text{L1}}(x;g+1)
  P_n^{\text{L}}(x;g+\ell)\n
  &\phantom{P_{\ell,n}^{\text{L1}}(x;\bm{\lambda})\eqdef}
  \ \ -\xi_{\ell-1}^{\text{L1}}(x;g+2)P_{n-1}^{\text{L}}(x;g+\ell),
  \label{PlnLag}\\
  &h_{\ell,n}^{\text{L1}}(\bm{\lambda})
  =\frac{n+g+2\ell-\frac12}{n+g+\ell-\frac12}\,h_n^{\text{L}}(g+\ell).
  \label{L1h}
\end{align}
\\[1pt]
{\bf trigonometric DPT}  undeformed ($\ell=0$) case:
\begin{align}
  &\bm{\lambda}^{\text{J}}\eqdef(g,h),
  \ \ \bm{\delta}^{\text{J}}=(1,1),\ \ g,h>0,
  \label{trigdat1}\\
  &\mathcal{E}_n^{\text{J}}(\bm{\lambda})\!=\!4n(n+g+h),
  \ \eta^{\text{J}}(x)\eqdef\cos2x,\ 0<x<\frac{\pi}{2},\!\!
  \label{trigdomain}\\
  &\phi_0^{\text{J}}(x;\bm{\lambda})\eqdef(\sin x)^g(\cos x)^h\n
  &\qquad\Leftrightarrow
  w_0^{\text{J}}(x;\bm{\lambda})\eqdef g\log\sin x+h\log\cos x,\\
  &P_n^\text{J}(x;\bm{\lambda})\eqdef P_n^{(g-\frac12,\,h-\frac12)}(x),\\
  &h_n^{\text{J}}(\bm{\lambda})=\frac{\Gamma(n+g+\frac12)\Gamma(n+h+\frac12)}
  {2\,n!\,(2n+g+h)\Gamma(n+g+h)}.
\end{align}
\\[1pt]
$\ell$-th deformed {\bf trigonometric DPT}:
\begin{align}
  &\xi_{\ell}^{\text{J1}}(x;\bm{\lambda})\eqdef
  P_{\ell}^{(-g-\ell-\frac12,\,h+\ell-\frac32)}(x),\quad h>g>0,
  \label{xiJ1}\\
  &P_{\ell,n}^{\text{J1}}(x;\bm{\lambda})\eqdef
  a_{\ell,n}^{\text{J1}}(x;\bm{\lambda})
  P_n^{\text{J}}(x;\bm{\lambda}+\ell\bm{\delta})\n
  &\phantom{P_{\ell,n}^{\text{J1}}(x;\bm{\lambda})=}
  +b_{\ell,n}^{\text{J1}}(x;\bm{\lambda})
  P_{n-1}^{\text{J}}(x;\bm{\lambda}+\ell\bm{\delta}),
  \label{Plnjac}\\
  &a_{\ell,n}^{\text{J1}}(x;\bm{\lambda})
  \eqdef\xi_{\ell}^{\text{J1}}(x;g+1,h+1)\n[-2pt]
  &\qquad
  +\frac{2n(-g+h+\ell-1)\,\xi_{\ell-1}^{\text{J1}}(x;g,h+2)}
  {(-g+h+2\ell-2)(g+h+2n+2\ell-1)}\n[-1pt]
  &\qquad
  -\frac{n(2h+4\ell-3)\,\xi_{\ell-2}^{\text{J1}}(x;g+1,h+3)}
  {(2g+2n+1)(-g+h+2\ell-2)},\\
  &b_{\ell,n}^{\text{J1}}(x;\bm{\lambda})\eqdef
  \frac{(-g+h+\ell-1)(2g+2n+2\ell-1)}{(2g+2n+1)(g+h+2n+2\ell-1)}\n
  &\phantom{b_{\ell,n}^{\text{J1}}(x;\bm{\lambda})\eqdef}
  \ \ \times\xi_{\ell-1}^{\text{J1}}(x;g,h+2),\\
  &h_{\ell,n}^{\text{J1}}(\bm{\lambda})
  =\frac{(n+g+\ell+\frac12)(n+h+2\ell-\frac12)}
  {(n+g+\frac12)(n+h+\ell-\frac12)}\n
  &\phantom{h_{\ell,n}^{\text{J1}}(\bm{\lambda})=}
  \ \ \times h_n^{\text{J}}(g+\ell,h+\ell).
  \label{bJ1}
\end{align}

For the proof of shape invariance \eqref{wlshapeinv}, see a recent paper
\cite{os17}.

%%%%%%%%%%%%%%%%%%%%%%%%%%%%%%%%%%%%%%%%%%%%%%%%%%%%%%%%%%%%%%%
%                                                             %
%  3. Another set of deformed oscillator potentials           %
%     \& $X_{\ell}$ Laguerre polynomials                      %
%                                                             %
%%%%%%%%%%%%%%%%%%%%%%%%%%%%%%%%%%%%%%%%%%%%%%%%%%%%%%%%%%%%%%%
\section{Another set of deformed oscillator potentials \& $X_{\ell}$
Laguerre polynomials}
\label{another}

\noindent
The {\em second set\/} of deformed radial oscillator and the
corresponding $X_{\ell}$ Laguerre polynomials are the following.
\\[2pt]
2-nd $\ell$-th deformed {\bf radial oscillator}:
\begin{align}
  &\xi_{\ell}^{\text{L2}}(x;\bm{\lambda})\eqdef
  L_{\ell}^{(-g-\ell-\tfrac{1}{2})}(x),\\
  &P_{\ell,n}^{\text{L2}}(x;\bm{\lambda})\n
  &\eqdef\Bigl(\xi_{\ell}^{\text{L2}}(x;g+1)
  -\frac{2n\,\xi_{\ell-2}^{\text{L2}}(x;g+1)}{2g+2n+1}\Bigl)
  P_n^{\text{L}}(x;g+\ell)\n
  &\qquad+\frac{2g+2n+2\ell-1}{2g+2n+1}\,\xi_{\ell-1}^{\text{L2}}(x;g)
  P_{n-1}^{\text{L}}(x;g+\ell),\\
  &h_{\ell,n}^{\text{L2}}(\bm{\lambda})=\frac{n+g+\ell+\frac12}{n+g+\frac12}
  \,h_n^{\text{L}}(g+\ell).
\end{align}
The action of the operators $\mathcal{A}_{\ell}(g)$ and
$\mathcal{A}_{\ell}(g)^{\dagger}$ on the eigenfunction $\phi_{\ell,n}(x;g)$
(or the action of the forward and backward shift operators on the
eigenpolynomial $P_{\ell,n}(\eta(x);g)$) is the same as the $\text{L1}$ case,
eq.(22) in \cite{os16}.
It is easy to verify that $\xi_{\ell}^{\text{L2}}(x;\bm{\lambda})$ is
of the same sign for $0<x<\infty$ so that the deformation is non-singular.
It is important to note that the first members of the deforming polynomials
are essentially the same for the $\text{L1}$ and $\text{L2}$:
\begin{equation}
  \xi_1^{\text{L1}}(x;g)=-\xi_1^{\text{L2}}(x;g).
\end{equation}
Since the normalisation of the $\xi$'s are irrelevant to the deformation,
the first deformed potential and the eigenpolynomials are identical:
\begin{equation}
  \mathcal{H}_1^{\text{L1}}(x;g)=\mathcal{H}_1^{\text{L2}}(x;g),\quad
  P_{1,n}^{\text{L1}}(x;g)=-P_{1,n}^{\text{L2}}(x;g).
\end{equation}
This together with the same situation in the Jacobi case \eqref{J12eq}
are one of the reasons why the second sets are not recognised earlier.
It is straightforward to verify that Quesne's type \Romannumeral{2}
potential eq.(2.17) lower sign of \cite{quesne2} is the same as the
second deformed potential $U_2^{\text{L2}}(x;\bm{\lambda})$, with the
replacements $\omega\to2$, $l\to g+1$.
Correspondingly her type \Romannumeral{2} $X_2$ Laguerre polynomials
$\tilde{L}_{2,\nu+2}^{(\alpha)}(z)$ eq.(2.26) agree with the above
$P_{\ell,n}^{\text{L2}}(x;g)$,
$\tilde{L}_{2,n+2}^{(g+\frac32)}(x)=2P_{2,n}^{\text{L2}}(x;g)$.
Her type \Romannumeral{1} polynomials also coincide with ours,
$\tilde{L}_{1,n+2}^{(g+\frac32)}(x)=2P_{2,n}^{\text{L1}}(x;g)$.

The above second set is obtained by a simple limiting procedure from
the deformed trigonometric DPT potential and the corresponding $X_{\ell}$
Jacobi polynomials \eqref{xiJ1}--\eqref{bJ1} given in the preceding section.

In fact, the limit formulas of the base polynomials
\begin{equation}
  \lim_{\beta\to\infty}P_n^{(\alpha,\,\pm\beta)}
  \bigl(1-2x\beta^{-1}\bigr)
  =L_n^{(\alpha)}(\pm x)
  \label{JtoL}
\end{equation}
are well known. The radial oscillator potential is known to be obtained
from the trigonometric DPT potential in the limit of infinite coupling
$h\to\infty$ with the rescaling of the coordinate:
\begin{equation}
  x=\frac{x^{\text{L}}}{\sqrt{h}},\quad
  0<x<\frac{\pi}{2}\Leftrightarrow 0<x^{\text{L}}<\frac{\pi}{2}\sqrt{h}\,.
  \label{rescale}
\end{equation}
We then have
\begin{align}
  &\eta^{\text{J}}(x)=1-2\eta^{\text{L}}(x^{\text{L}})h^{-1}+O(h^{-2}),\\
  &\lim_{h\to\infty}\bigl(w_0^{\text{J}}(x;g,h)+\tfrac12g\log h\bigr)
  =w_0^{\text{L}}(x^{\text{L}};g).
\end{align}
Since a constant shift of the prepotential does not affect the Hamiltonian,
these lead to the limit relations for the Hamiltonians and eigenfunctions
\begin{align}
  &\lim_{h\to\infty}P_n^{\text{J}}\bigl(\eta^{\text{J}}(x);g,h\bigr)
  =P_n^{\text{L}}\bigl(\eta^{\text{L}}(x^{\text{L}});g\bigr),\\
  &\lim_{h\to\infty}h^{-1}\mathcal{H}_0^{\text{J}}(x;g,h)
  =\mathcal{H}_0^{\text{L}}(x^{\text{L}};g),\\
  &\lim_{h\to\infty}h^{-1}\mathcal{E}_n^{\text{J}}(g,h)
  =\mathcal{E}_n^{\text{L}}(g).
\end{align}
Similar limit formulas hold for the $\ell$ deformed systems:
\begin{align}
  &\lim_{h\to\infty}\xi_{\ell}^{\text{J1}}\bigl(\eta^{\text{J}}(x);g,h\bigr)
  =\xi_{\ell}^{\text{L2}}\bigl(\eta^{\text{L}}(x^{\text{L}});g\bigr),\\
  &\lim_{h\to\infty}\bigl(w_{\ell}^{\text{J1}}(x;g,h)
  +\tfrac12(g+\ell)\log h\bigr)
  =w_{\ell}^{\text{L2}}(x^{\text{L}};g),\\
  &\lim_{h\to\infty}h^{-1}\mathcal{H}_{\ell}^{\text{J1}}(x;g,h)
  =\mathcal{H}_{\ell}^{\text{L2}}(x^{\text{L}};g),\\
  &\lim_{h\to\infty}P_{\ell,n}^{\text{J1}}\bigl(\eta^{\text{J}}(x);g,h\bigr)
  =P_{\ell,n}^{\text{L2}}\bigl(\eta^{\text{L}}(x^{\text{L}});g\bigr).
\end{align}

By changing the roles of $g$ and $h$ in \eqref{xiJ1}--\eqref{bJ1}, we obtain
the {\em second set\/} of deformed trigonometric DPT.
\\[2pt]
2-nd $\ell$-th deformed {\bf trigonometric DPT}:
\begin{align}
  &\xi_{\ell}^{\text{J2}}(x;\bm{\lambda})\eqdef
  P_{\ell}^{(g+\ell-\frac32,\,-h-\ell-\frac12)}(x),\quad g>h>0,
  \label{J2xi}\\
  &P_{\ell,n}^{\text{J2}}(x;\bm{\lambda})\eqdef
  a_{\ell,n}^{\text{J2}}(x;\bm{\lambda})
  P_n^{\text{J}}(x;\bm{\lambda}+\ell\bm{\delta})\n
  &\phantom{P_{\ell,n}^{\text{J2}}(x;\bm{\lambda})=}
  +b_{\ell,n}^{\text{J2}}(x;\bm{\lambda})
  P_{n-1}^{\text{J}}(x;\bm{\lambda}+\ell\bm{\delta}),\\
  &a_{\ell,n}^{\text{J2}}(x;\bm{\lambda})\eqdef
  \xi_{\ell}^{\text{J2}}(x;g+1,h+1)\n[-2pt]
  &\qquad
  -\frac{2n(g-h+\ell-1)\,\xi_{\ell-1}^{\text{J2}}(x;g+2,h)}
  {(g-h+2\ell-2)(g+h+2n+2\ell-1)}\n[-1pt]
  &\qquad
  -\frac{n(2g+4\ell-3)\,\xi_{\ell-2}^{\text{J2}}(x;g+3,h+1)}
  {(2h+2n+1)(g-h+2\ell-2)},\\
  &b_{\ell,n}^{\text{J2}}(x;\bm{\lambda})\eqdef
  \frac{(g-h+\ell-1)(2h+2n+2\ell-1)}{(2h+2n+1)(g+h+2n+2\ell-1)}\n
  &\phantom{b_{\ell,n}^{\text{J2}}(x;\bm{\lambda})\eqdef}
  \ \ \times\xi_{\ell-1}^{\text{J2}}(x;g+2,h),\\
  &h_{\ell,n}^{\text{J2}}(\bm{\lambda})
  =\frac{(n+h+\ell+\frac12)(n+g+2\ell-\frac12)}
  {(n+h+\frac12)(n+g+\ell-\frac12)}\n
  &\phantom{h_{\ell,n}^{\text{J2}}(\bm{\lambda})=}
  \ \ \times h_n^{\text{J}}(g+\ell,h+\ell).
  \label{J2h}
\end{align}
The action of the operators $\mathcal{A}_{\ell}(\bm{\lambda})$ and
$\mathcal{A}_{\ell}(\bm{\lambda})^{\dagger}$ on the eigenfunction
$\phi_{\ell,n}(x;\bm{\lambda})$ (or the action of the forward and
backward shift operators on the eigenpolynomial
$P_{\ell,n}(\eta(x);\bm{\lambda})$) is the same as the $\text{J1}$ case,
eq.(37) in \cite{os16}.
Again the deforming polynomial $\xi_{\ell}^{\text{J2}}(x;\bm{\lambda})$
($g>h>0$) is of the same sign in $-1<x<1$ and the deformation is
non-singular.
It is interesting to note that the first members of the deforming
polynomials are essentially the same for the J1 and J2:
\begin{equation}
  \xi_1^{\text{J1}}(x;g,h)=-\xi_1^{\text{J2}}(x;g,h).
  \label{J12eq}
\end{equation}
This means that there is no difference between the J1 and J2 deformations
in the $\ell=1$ case.

In fact this is not a new set but simply a `mirror image' of the first
set \eqref{xiJ1}--\eqref{bJ1}, under the transformation:
\begin{align}
  x\to y\eqdef\frac{\pi}{2}-x,
  \quad 0<x<\frac{\pi}{2},\quad 0<y<\frac{\pi}{2},
\end{align}
and renaming of the coupling constants $g\leftrightarrow h$.
By the parity property of the Jacobi polynomial
$P_n^{(\alpha,\,\beta)}(-x)=(-1)^nP_n^{(\beta,\,\alpha)}(x)$ and
\begin{equation}
  \eta^{\text{J}}(x)=-\eta^{\text{J}}(y),
  \quad w_0^{\text{J}}(x;g,h)=w_0^{\text{J}}(y;h,g),
\end{equation}
we obtain
\begin{equation}
  \xi_{\ell}^{\text{J2}}(x;g,h)
  =(-1)^{\ell}\xi_{\ell}^{\text{J1}}(-x;h,g).
\end{equation}
Hence the above assertion is demonstrated:
\begin{align}
  &\mathcal{H}_{\ell}^{\text{J2}}(x;g,h)
  =\mathcal{H}_{\ell}^{\text{J1}}\bigl(y;h,g\bigr),\\
  &P_{\ell,n}^{\text{J2}}(x;g,h)
  =(-1)^{\ell+n}P_{\ell,n}^{\text{J1}}(-x;h,g).
  \label{plnJ2=plnJ1}
\end{align}
Now it is easy to verify that the original (L1) set of the $\ell$-th
deformed oscillator and the $X_{\ell}$ Laguerre polynomials
\eqref{L1xi}--\eqref{L1h} are obtained from the second set of the
trigonometric DPT and the $X_{\ell}$ Jacobi polynomials
\eqref{J2xi}--\eqref{J2h} in the same $h\to\infty$ limit with the
corresponding rescaling of the coordinate \eqref{rescale}:
\begin{align}
  &\lim_{h\to\infty}\xi_{\ell}^{\text{J2}}\bigl(\eta^{\text{J}}(x);g,h\bigr)
  =\xi_{\ell}^{\text{L1}}\bigl(\eta^{\text{L}}(x^{\text{L}});g\bigr),\\
  &\lim_{h\to\infty}\bigl(w_{\ell}^{\text{J2}}(x;g,h)
  +\tfrac12(g+\ell)\log h\bigr)
  =w_{\ell}^{\text{L1}}(x^{\text{L}};g),\\
  &\lim_{h\to\infty}h^{-1}\mathcal{H}_{\ell}^{\text{J2}}(x;g,h)
  =\mathcal{H}_{\ell}^{\text{L1}}(x^{\text{L}};g),\\
  &\lim_{h\to\infty}P_{\ell,n}^{\text{J2}}\bigl(\eta^{\text{J}}(x);g,h\bigr)
  =P_{\ell,n}^{\text{L1}}\bigl(\eta^{\text{L}}(x^{\text{L}});g\bigr).
\end{align}

%%%%%%%%%%%%%%%%%%%%%%%%%%%%%%%%%%%%%%%%%%%%%%%%%%%%%%%%%%%%%%%
%                                                             %
%  4. Summary and Comments                                    %
%                                                             %
%%%%%%%%%%%%%%%%%%%%%%%%%%%%%%%%%%%%%%%%%%%%%%%%%%%%%%%%%%%%%%%
\section{Summary and Comments}

We have presented a new set of infinitely many deformed radial oscillator
potentials, which are shape invariant and thus exactly solvable.
The corresponding $X_{\ell}^{\text{L2}}$ Laguerre polynomials are also
given. They can be derived from the first set of trigonometric DPT and
the corresponding $X_{\ell}$ Jacobi polynomials in a certain limit.
The shape invariance relations for the first and second sets of
deformations are attributed to the same cubic identities \cite{os18},
both for the Laguerre and Jacobi cases. Various properties of the $X_{\ell}$
Laguerre and Jacobi polynomials of both kinds will be discussed in a
forthcoming article \cite{hos}. In particular, we will present equivalent
but much simpler forms of the $X_{\ell}$ Laguerre and Jacobi polynomials.
It would be a good challenge to search matrix models associated with these
exceptional orthogonal polynomials.

In Quesne's paper \cite{quesne2} the explicit forms are given of the
candidates of the second type of the exceptional ($X_2$) Laguerre
polynomials together with the corresponding deformed potential, which
are shown to be the same as those given in this Letter.
In the same paper \cite{quesne2}, Quesne reported two non shape invariant
but exactly solvable potentials called type \Romannumeral{3}, each related
to the radial oscillator and the DPT.
Let us remark that they can be easily obtained by applying Adler's
\cite{adler} modification
of Crum's method \cite{crum} to the radial oscillator and DPT, with
the specification to remove the first and second excited states of the
undeformed system. The resulting eigenfunctions (the groundstate included)
constitute the complete set of the Hilbert space, as proved in the paper
\cite{adler}.
Let us note that starting from an exactly solvable system, an infinite
variety of exactly solvable  potentials and the corresponding
eigenfunctions can be constructed by Adler's method \cite{adler}.
None of the derived systems, however, is shape invariant even if the
starting system is.

Before closing this Letter, let us also mention that the deformation
in terms of a degree $\ell$ eigenpolynomial, applied to the discrete
quantum mechanical Hamiltonians for the Wilson and Askey-Wilson
polynomials, produced two sets of infinitely many shape invariant
systems together with exceptional ($X_{\ell}$) Wilson and Askey-Wilson
polynomials ($\ell=1,2,\ldots$) \cite{os17}.

\bigskip
%%%%%%%%%%%%%%%%%%%%%%%%%%%%%%%%%%%%%%%%%%%%%%%%%%%%%%%%%%%%%%%
%                                                             %
%  Acknowledgments                                            %
%                                                             %
%%%%%%%%%%%%%%%%%%%%%%%%%%%%%%%%%%%%%%%%%%%%%%%%%%%%%%%%%%%%%%%
This work is supported in part by Grants-in-Aid for Scientific Research
from the Ministry of Education, Culture, Sports, Science and Technology,
No.19540179.

%%%%%%%%%%%%%%%%%%%%%%%%%%%%%%%%%%%%%%%%%%%%%%%%%%%%%%%%%%%%%%%
%                                                             %
%  References                                                 %
%                                                             %
%%%%%%%%%%%%%%%%%%%%%%%%%%%%%%%%%%%%%%%%%%%%%%%%%%%%%%%%%%%%%%%


\begin{thebibliography}{99}
%
% for zero-space: \hspace{0pt}

\bibitem{bochner}
S.\,Bochner,
%``\"Uber Sturm-Liouvillesche Polynomsysteme,"
%Math. Zeit. {\bf 29} (1929) 730-736.
Math. Zeit. {\bf 29} (1929) 730.

\bibitem{gomez}
D.\,G\'omez-Ullate, N.\,Kamran and R.\,Milson,
%``An extension of Bochner's problem: exceptional invariant subspaces,''
{\tt arXiv:0805.3376\hspace{0pt}[math-ph]};
%``An extended class of orthogonal polynomials defined by a
%Sturm-Liouville problem,''
J. Math. Anal. Appl. {\bf 359} (2009) 352,
{\tt arXiv:0807.\hspace{0pt}3939[math-ph]};
%``Supersymmetry and algebraic Darboux transformations,"
%J. Phys. {\bf A37} (2004) 10065-10078.
J. Phys. {\bf A37} (2004) 10065.

\bibitem{quesne}
C.\,Quesne,
%``Exceptional orthogonal polynomials, exactly solvable potentials
%and supersymmetry,''
%J. Phys. {\bf A41} (2008) 392001 (6 pages),
J. Phys. {\bf A41} (2008) 392001,
{\tt arXiv:0807.4087\hspace{0pt}[quant-ph]};
B.\,Bagchi, C.\,Quesne and R.\,Roychoudhury,
%``Isospectrality of conventional and new extended potentials,
%second-order supersymmetry and role of PT symmetry,"
%Pramana J. Phys. {\bf 73} (2009) 337-347,
Pramana J. Phys. {\bf 73} (2009) 337,
{\tt arXiv:0812.1488[quant-ph]}.

\bibitem{genden}
L.\,E.\,Gendenshtein,
%``Derivation of exact spectra of the Schrodinger equation by means of
%supersymmetry,''
%JETP Lett. {\bf 38} (1983) 356-359.
JETP Lett. {\bf 38} (1983) 356.

\bibitem{os16}
S.\,Odake and R.\,Sasaki,
%``Infinitely many shape invariant potentials and new orthogonal polynomials,'' 
%Phys. Lett. {\bf B679} (2009) 414-417,
Phys. Lett. {\bf B679} (2009) 414,
{\tt arXiv:\hspace{0pt}0906.0142[math-ph]}.

\bibitem{os18}
S.\,Odake and R.\,Sasaki,
%``Infinitely many shape invariant potentials and
%cubic identities of the Laguerre and Jacobi polynomials,''
{\tt arXiv:0911.1585[math-ph]}.

\bibitem{quesne2}
C.\,Quesne,
%``Solvable Rational Potentials and Exceptional Orthogonal Polynomials
%in Supersymmetric Quantum Mechanics,"
%SIGMA {\bf 5} (2009) 084 (24 pages),
SIGMA {\bf 5} (2009) 084,
{\tt arXiv:0906.2331[math-ph]}.

\bibitem{tanaka}
T.\,Tanaka,
%``$\mathcal{N}$-fold Supersymmetry and Quasi-solvability Associated with
%$X_2$-Laguerre Polynomials,''
{\tt arXive:0910.0328[math-ph]}.

\bibitem{hos}
C-L.\,Ho, S.\,Odake and R.\,Sasaki,
``Properties of the exceptional ($X_{\ell}$) Laguerre and Jacobi polynomials,"
YITP-09-70, in preparation.

\bibitem{adler}
V.\,\'E.\, Adler,
%``A Modification of Crum's Method,''
%Theo. Math. Phys. {\bf 101} (1994) 1381-1386.
Theor. Math. Phys. {\bf 101} (1994) 1381.

\bibitem{crum}
M.\,M.\,Crum,
%``Associated Sturm-Liouville systems,''
Quart. J. Math. Oxford Ser. (2)
%{\bf 6} (1955) 121-127,
{\bf 6} (1955) 121,
{\tt arXiv:physics/9908019}.

\bibitem{os17}
S.\,Odake and R.\,Sasaki,
%``Infinitely many shape invariant discrete quantum mechanical
%systems and new exceptional orthogonal polynomials related to the
%Wilson and Askey-Wilson polynomials,"
%Phys. Lett. {\bf B682} (2009) 130-136,
Phys. Lett. {\bf B682} (2009) 130,
{\tt arXiv:\hspace{0pt}0909.3668[math-ph]}.

\end{thebibliography}
\end{document}